\documentclass[12pt, a4paper]{article}
\usepackage[font=footnotesize]{caption}
\usepackage[utf8]{inputenc}
\usepackage{graphicx}
\usepackage{authblk}
\usepackage{booktabs}
\usepackage{graphicx}
\usepackage{verbatim}
\usepackage{url}
\usepackage{caption}
\usepackage{subcaption}
\usepackage{indentfirst}
\usepackage{url}
\title{Enzyme Similarity Networks}
\author{Renan dos Reis  and  Luciano da F.  Costa}

\date{
S\~ao Carlos Institute of Physics - DFCM \protect\\
University of S\~ao Paulo \protect\\
P.O.  Box 369, S\~ao Carlos, S.P.  \protect\\
13560-970 Brazil \\ \vspace{0.5cm}
\emph{4th April, 2022}
}

\begin{document}
\maketitle

\begin{abstract}
There is a crescent use of enzymes in multiple industries and sciences, ranging from materials and fuel synthesis to pharmaceutical and food production.  Their applicability in this variety of fields depends not only on their biochemical function but also on their physicochemical properties.  In the present work, we describe how the coincidence methodology can be employed to construct similarity networks of seventy well-studied enzymes of the Glycoside Hydrolase Family 13 and to identify communities of physicochemically related enzymes.  More specifically, each of the selected enzymes is mapped into a network node, while the links between pairs of enzymes are determined by the coincidence similarity between selected physicochemical features of interest.  The obtained networks have modularity and number of isolated nodes optimized respectively to two parameters involved in the coincidence methodology, resulting in highly modular networks.  In order to investigate the effect of the considered physicochemical features on the enzymes relationships, the coincidence-based method also is applied to create a meta-network, in which the enzymes similarity networks obtained by the combination of every possible feature becomes nodes of a feature combination network, and the coincidence similarity between those networks defines the respective links.  The obtained feature combination network systematically and comprehensively indicates the impact of the selected physicochemical features on enzyme similarity.  Several interesting results are reported and discussed, including the identification of subgroups of enzymes with similar physicochemical features within catalytical classes, providing important information for the selection and design of enzymes for targeted biotechnological applications.
\end{abstract}


\section{Introduction}

Several of the most complex real-world systems can be found in biology, extending from micro to macroscopic time and spatial scales.  Examples at micro and mesoscopic level include the immune system, the basal metabolism, and the nervous system.  At more macroscopic levels, we have food webs and ecology.

A great deal of the biological complexity stems not only from the vast number of elements that are typically interacting, but also from the diversity of types of these entities.  Of exceptional scientific relevance are the systems that involves proteins, which appear in biological systems in an impressive diversity of sizes and types, with multiple functionalities and physicochemical peculiarities.

Proteins are polymers of amino acids, each chosen among 22 possible types.  Each amino acid has specific electrical, chemical, and other physical properties that dictate how they behave and how the protein folds in 3D space.  The protein and all its resulting features on the 3D space -- including but not limited to its geometry, topology, volume, stability, and solubility -- will depend on its amino acid composition and will enable its biochemical function~\cite{Creighton2010}.  Thus, the vast number of possibilities to assemble amino acid chains is directly connected to the numerous possible activities that proteins can perform~\cite{Creighton2010}.

An activity of particular importance for proteins is their capacity to act as catalysts.  This activity is crucial for the existence of life, as almost all biochemical reactions in living systems depend on proteins to occur at the rates needed to maintain biological systems~\cite{Creighton2010}.  This role is so important that proteins that act as catalysts are specially called \emph{enzymes}.  

There are four perspectives from which the catalytic activity of enzymes can be better appreciated.  The first is that enzymes increase the rates of very specific chemical reactions by selectively interacting with substrates, stabilizing transition states, and releasing chemical products~\cite{Creighton2010}.  The second is that the same enzyme can have more than one catalytic activity.  For instance, the neopullulanase enzyme of \textit{Geobacillus stearothermophilus} can participate in both the converting of pullulan in panose and the converting of cyclomaltodextrin in linear maltodextrin~\cite{Hondoh2003}.  The third perspective is that the same enzyme type can exist in several distinct organisms, extending along with both related and unrelated phylogenetic levels.  For example, pullulanase enzymes can be found in \textit{Bacillus subtilisand} and \textit{Thermococcus kodakarensis}, two bacteria species, as well as in \textit{Oryza sativa} and \textit{Spinacia oleracea}, two plant species~\cite{CAZY}.  The fourth perspective is that enzymes can be applied outside of their source organism to catalyze reactions of biotechnological interest~\cite{Buchholz2012}.

The catalytic activity of an enzyme stems from two principal aspects: (i) its 3D geometric shape; and (ii) the physical and chemical properties along its chain of amino acids~\cite{Creighton2010}.The 3D geometric shape is very important, as the interactions between the enzyme and its substrates occur at specific active sites in the enzyme structure~\cite{Creighton2010}.  For instance, the active site of an enzyme can be a concave and positively charged cavity that can interact with a convex and negatively charged substrate, making it stable and prone to a chemical reaction such as hydrolysis.  However, although there are multiple computational and experimental techniques to study enzyme structure, this still is an arduous task.  

For this reason, the study of the activity of an enzyme and its possible applications in biotechnological systems usually begins with the characterization of its physical and chemical properties~\cite{Buchholz2012}.  Subsequent analysis involves organizing a set of enzymes of particular interest according to their mutual physicochemical similarity.  This is of particular interest because enzymes with the same catalytic activity and with similar physicochemical properties can be synthesized, purified, and applied in similar conditions, thus being interchangeable in their biotechnological applications~\cite{Buchholz2012}.  

Over the last decade, multiple efforts have been carried out to classify enzymes based on their functional characteristics.  The main one is the creation of the Enzyme Commission (EC) number system by the International Union of Biochemistry and Molecular Biology (IUBMB)~\cite{IUBMB}.  This commission labels enzymes with a four-part EC number based on their catalytic activities.  For example, the neopullulanase enzyme of \textit{Geobacillus stearothermophilus} mentioned above receives the EC numbers 3.2.1.135 for its neopullulanase activity and 3.2.1.54 for its cyclomaltodextrinase~\cite{CAZY}.  These four sequential numbers have different meanings: the first indicates the main catalytic class of the catalysis (e.g., 3 is for hydrolases); the second is for the subclass based on the target chemical structure (e.g., 3.2 is for glycosylases - targets glycosyl groups); the third is for further information on the chemical reaction (e.g., 3.2.1 is for glycosidases - targets O- and S-glycosyl groups); and the fourth is a serial number for the specific reaction~\cite{Buchholz2012,McDonald2009}.  This classification method is very convenient for scientifical and biotechnological purposes, as enzymes are promptly grouped with other proteins with the same biochemical applications, irrespective of their original species.  The EC numbers system can be easily accessed on the ExplorEnz database, available at https://www.enzyme-database.org/~\cite{McDonald2009}.

Other important efforts on curating and classifying proteins are: the BRENDA database (https://www.brenda-enzymes.org), with nomenclature, functional, kinetic, organism-related, structural and experimental information on enzymes~\cite{BRENDA}; the BioCatNet database system (https://biocatnet.de/), focused on catalytic and amino acid sequence information of enzymes for engineering purposes~\cite{BioCatNet}; the carbohydrate-active enzymes database (CAZy) (http://www.cazy.org/), specialized on the organization of genomic, structural and biochemical features of carbohydrate-active enzymes (CAZymes), as well as their classification into protein families~\cite{CAZY}.  These resources can assist biotechnology researchers to both prospect and engineer enzymes with better features for their intended application~\cite{Fasim2021}.  

The knowledge available on those databases could be better explored by identifying communities of enzymes with similar properties within the groups assembled through phylogenetic, structural and catalytic analysis.  For this purpose, in the present work, we apply the coincidence methodology~\cite{CostaCCompl} to obtain enzyme similarity networks based on physicochemical data.  The coincidence similarity was developed as an extension of the Jaccard similarity index capable of being applied to real values~\cite{CostaJaccard,CostaSimilarity} and able to incorporate information about the relative interiority (or overlap, e.g..g.~\cite{Kavitha}) between two datasets.  When compared to more traditional similarity indices as the cosine similarity and correlation, the coincidence index has been found to allow more strict quantification of the pairwise similarity between two datasets~\cite{CostaSimilarity,CostaComparing}.

The coincidence methodology has been proposed~\cite{CostaCCompl} as a means to translate datasets of entities characterized by respective features into networks presenting enhanced connectivity details and modularity.  It has been applied with encouraging success to several types of data~\cite{CostaCaleidoscope,Costa_ElementaryParticles} and interesting problems including city characterization~\cite{Costa_CitiesSimilarity} and motifs identification~\cite{Domingues_CityMotifs}.

In the present work, we apply the coincidence methodology to translate seventy well-studied enzymes of biotechnological interest into similarity networks.  Each node of the similarity network corresponds to an enzyme, and each connection between pairs of nodes are specified by the pairwise coincidence similarity between their respective physicochemical features -- namely number of amino acids, molecular weight, theoretical isoelectric point (pI), instability index, aliphatic index, and grand average of hydropathicity (GRAVY)~\cite{ProtParam}.  

The translation procedure involves two parameters, namely $\alpha$ and $T$, respectively controlling the relative contribution of negative and positive feature values and the threshold for removing weaker connections~\cite{CostaCCompl}.  Different properties of the similarity networks can be optimized while systematically varying $\alpha$ and $T$~\cite{CostaCCompl,Domingues_CityMotifs}.  In order to find subgroups of connected entities using similarity networks, one important property that can be optimized is the modularity.  However, often the created network with the parameter configuration leading to maximum modularity is also characterized by a large number of isolated nodes.  In order to identify the parameter configurations leading to high (but not necessarily optimal) modularity while simultaneously minimizing the number of isolated nodes, we adopted a supervised combined optimization index reflecting both these sought properties.  

The enzymes selected for this work are CAZymes of the glycoside hydrolase family 13 (GH13)~\cite{CAZY}.  They cover four types of catalytic activities~\cite{CAZY}, each with particular biotechnological interests: hexosyltransferases (EC 2.4.1.-), applied in the production of sugars with novel properties, both in material and food sciences~\cite{biotech241a,biotech241b,biotech241c}; phosphate $\alpha$-maltosyltransferases (EC 2.4.99.16), enzymes of potential pharmaceutical interest~\cite{biotech2499241,biotech2499a,biotech2499b};  glycosidases, enzymes that hydrolyse O- and S-glycosyl compounds (EC 3.2.1.-), applied in starch and baking industries~\cite{biotech321}; and intramolecular transferases of maltodextrin, sucrose or maltose (EC 5.4.99.-), used for the production of valuable sugars, such as trehalose~\cite{biotech5499}.  In the translation of these enzymes into similarity networks, we calculated the combined optimization index based on the EC number of each enzyme, which was tuned by controlling the $\alpha$ and $T$ parameters.  The proposed method led to enzyme similarity networks with both a small number of isolated nodes and high levels of catalytic similarity within the obtained clusters.

The results typically obtained in supervised classification are known to depend critically on the set of features adopted to characterize the data elements of interest (i.e.~individual enzymes).  In the present work, we employ the feature analysis methodology described in~\cite{Costa_ElementaryParticles} as the means for studying, in a systematic and complete manner, the contribution of each of the considered physicochemical features on the obtained enzyme similarity networks.  More specifically, each of the possible combinations of physicochemical features is considered for obtaining optimal enzyme similarity networks, and the similarity between each of these networks is then gauged by using the coincidence index, resulting in a new network that expresses in a comprehensive and systematic manner the effects of considering different sets of features on the construction of enzyme similarity networks.  Therefore, one of the contributions of the approach reported in this work is the inspection of how the adopted physicochemical features, calculated based on the properties of each amino acid that forms the enzymes, can for groups of enzymes within and among the four original types of catalytic activities.  In this sense, a good discrimination between enzymes of the four original types obtained by using the coincidence approach would indicate, among other implications, that these measurements can be used as subsidies for further characterizing the physicochemical space of each enzyme classes.

Several relevant contributions and results are reported and discussed in this work.  Combined with the potential of the coincidence methodology for providing more strict quantification of the similarity between two mathematical structures, the use of the composed optimization index favoring simultaneous maximization of the modularity and minimization of the number of isolated nodes yielded markedly uniform groups of connected enzymes (respectively to their EC number) while minimizing the number of isolated enzymes.  In this work, this is achieved by controlling the contribution of the features with the same or opposite signs on the resulting similarity as provided by the $\alpha$ parameter.  Remarkably, the optimal configuration could not have been achieved using the standard coincidence formula without resorting to the $\alpha$ parameter.  These results provide an important subsidy for the selection and design of enzymes for biotechnological applications based on both their catalytic activity and physicochemical profile.

The current work begins by presenting the adopted enzyme data and basic concepts and methods to be applied, including the coincidence similarity index, network modularity and its optimization, as well as the methodology for systematic analysis of the effect of the adopted features on the obtained protein similarity networks.

\section{Materials and Methods}

In this section are presented the model enzymes under study, their respective physicochemical features, the coincidence similarity and its application to translated datasets into networks, the concept of network modularity and its optimization, as well as the description of the systematic methodology for studying the effect of the adopted features on the obtained enzyme similarity networks.

\subsection{The Model Enzymes}

The seventy enzymes considered for this work were selected from the CAZy database, which has been used as a tool for the study of CAZymes since 1998~\cite{CAZY}.  This database categorizes CAZymes into five classes, each class with several families of structurally-related proteins~\cite{CAZY}.  One family of great biotechnological interest and with extensive experimental data available is the GH13, of the glycoside hydrolases class, composed of an evolutionary diversified set of enzymes that hydrolyze glycosidic bonds~\cite{GH13}.

The CAZy database has 129333 assigned enzymes under the GH13 family, from which only 147 are both characterized and have their structure available~\cite{CAZY}.  In this work, we will refer to these 147 enzymes as "well-studied".

Furthermore, the GH13 family has 30 different assigned EC numbers~\cite{CAZY}.  To simplify this collection of catalytic activities, we shorten the EC number to its third digit (e.g., the EC number 3.2.1.1 becomes 3.2.1), thus keeping enzymes with similar catalytic activities in the same group while increasing the size of those groups.  In this manner, all of the GH13 enzymes are distributed into four \emph{enzyme types}: 2.4.1, 2.4.99, 3.4.1,  5.4.99.  The description of the enzyme types is summarized in Table \ref{tab:ec}.

Among the 147 well-studied enzymes, we selected 70 model enzymes based on their evolutionary diversity, reported cases of biotechnological interest, and quality of the data available on CAZy~\cite{CAZY} and UniProt~\cite{UNIPROT} databases.  Additional information on the adopted model enzymes is in Supplementary Material 1.  The distribution of the model enzymes into the four enzyme types, the description of their catalytic activity, and their potential biotechnological uses are presented in Table \ref{tab:ec}.  

\begin{table}[h!]
\caption{Distribution of the 70 model enzymes into the four types of enzymes under study.  The model enzymes were selected from a pool of 147 well-studied CAZymes from the GH13 family.  Each enzyme type corresponds to a simplified EC number relative to a set of similar catalytic activities.  The asterisk indicates that all the well-studied enzymes available of that specific type have been used in this work.  \label{tab:ec}}
\resizebox{\textwidth}{!}{
\begin{tabular}{c|ccc} \toprule
Simplified EC & Description & \begin{tabular}{@{}c@{}}Number of \\ Model Enzymes \end{tabular} & \begin{tabular}{@{}c@{}}Biotechnological \\ Interest/Uses \end{tabular}  \\ \midrule
2.4.1 & Hexosyltransferases & 18 & \begin{tabular}{@{}c@{}}Production of sugars \\ with novel properties~\cite{biotech241a,biotech241b,biotech241c}\end{tabular} \\
\\
2.4.99 & \begin{tabular}{@{}c@{}}Transferring glycosil groups other than \\ hexosyl, pentosyl and sialyl \end{tabular} & 3* & Pharmaceutical interest~\cite{biotech2499241,biotech2499a,biotech2499b} \\
\\
3.4.1 & Glycosidases & 38 & Starch and baking industries~\cite{biotech321}\\
\\
5.4.99 & \begin{tabular}{@{}c@{}}Intramolecular transferases of \\ groups other than acyl, phospho, \\ amyno and hydroxy groups \end{tabular}
  & 11* & \begin{tabular}{@{}c@{}}Production of valuable sugars, \\  such as trehalose~\cite{biotech5499}\end{tabular}\\ \bottomrule
\end{tabular}
}
\end{table}

We collected the amino acid sequences of all chosen model enzymes in Uniprot~\cite{UNIPROT} by using the accession numbers provided by CAZy~\cite{CAZY}.  The amino acid sequences were used to calculate six physicochemical features important for the synthesis, purification and application of enzymes, namely:

\begin{enumerate}
  \item Number of amino acids -- It is one of the most fundamental features of a protein.  Since it is directly proportional to the length of its respective gene, this feature plays a role in assembling the right expression system, which can have restrictions based on gene length~\cite{Bajpai2014}.  Furthermore, the number of amino acids is important for size exclusion chromatography, a technique that separates proteins based on their capability to infiltrate and be retained in a porous gel medium, which depends on protein size and weight~\cite{Wilson2010,Buchholz2012}.  The target proteins can be purified by selecting a gel with compatible pore size, thus the purification of enzymes with similar sizes via molecular exclusion chromatography can be performed using the same chromatography equipment.
  
  \item Molecular weight -- Just like the previous, this is an important feature of size-exclusion chromatography~\cite{Wilson2010,Buchholz2012}.  Furthermore, the molecular weight can be exploited in ultrafiltration, a size-exclusion technique that consists in passing the solution with the target protein through a semipermeable membrane that retains molecules with a minimal molecular weight~\cite{Buchholz2012}.  Additionally, enzymes can be retained in semipermeable membrane to be used in membrane reactors~\cite{Buchholz2012}.  Therefore, enzymes with similar molecular weights could be purified or used on membrane reactors using the same semipermeable membranes.
  
  \item Theoretical pI -- The pI of an enzyme is the pH when its net charge is 0.  This property is important for setting the experimental conditions of ion-exchange chromatography.  This type of chromatography is used to separate and purify a target enzyme. It consists in passing the solution that contains the desired enzyme in a matrix with selected ion-exchange groups~\cite{Wilson2010,Buchholz2012}.  If the target enzyme is in a solution with a pH above its pI, it will have a negative surface charge and will be trapped in a matrix with a positively charged ion-exchange group (anion-exchange chromatography), while if the target enzyme is in a solution with a pH below its pI, it will have a positive surface charge and will be trapped in a matrix with negatively charged ion-exchange group (cation-exchange chromatography)~\cite{Wilson2010,Buchholz2012}.  Therefore, proteins with similar pI could be purified from a solution using a similar chromatography apparatus.  Such principle is also used in bifunctional or mixed-mode chromatography~\cite{Buchholz2012}
  
  \item Instability index -- Is a fundamental property used to predict the stability of the enzyme \textit{in vitro}~\cite{Guruprasad1990,ProtParam}.  Enzymes with an instability index lower than 40 are considered stable~\cite{Guruprasad1990}, and lower values of this index are expected to be related to more stable proteins.  Enzymes with similar instability index should demand similar experimental and industrial treatments during their purification and application -- e.g., suitable temperature to perform purification processes, possibility to perform denaturation fractionation, use of protease inhibitors in solution to avoid the destruction of the desired enzyme~\cite{Burgess2009}.

  \item Aliphatic index -- This feature corresponds to the relative volume occupied by aliphatic side chains, and it indirectly indicates the thermostability of the enzyme ~\cite{ProtParam,AliphaticIndex}.  Similar to the previous index, enzymes with the same aliphatic index should have similar responses to temperature, thus being an important characteristic to evaluate the suitability of an enzyme for industrial applications.
  
  \item Grand average of hydropathicity (GRAVY) -- It indicates the average hydropathicity, i.e. the hydrophilic or hydrophobic behavior, of the amino acids that composes the enzyme~\cite{ProtParam,GRAVY1}.  A positive GRAVY value indicates that the protein is hydrophobic and a negative value that it is hydrophilic~\cite{ProtParam,GRAVY1}.  Enzymes with similar GRAVY values will have a similar behavior toward the water, thus enabling the use of the same technical apparatus for purification through hydrophobic/hydrophilic interaction chromatography~\cite{Buchholz2012,Wilson2010,GRAVY2}.  Additionally, enzymes with similar GRAVY will have similar solubility to different solvents, therefore increasing the possibility to apply either of them in the same solution conditions.
\end{enumerate}

The calculations were performed with the ProtParam tool, available on the ExPASy server (https://web.expasy.org/protparam/)~\cite{ProtParam}.  Although several experimental methods are available to estimate the physicochemical properties of enzymes, this procedure was used to standardize the data acquisition for this large dataset.

\subsection{The Coincidence Similarity}

Quantifying the pairwise similarity between two mathematical structures represents one of the most essential operations in science and technology.  While the Jaccard~\cite{Jaccard} and other related indices (e.g.~\cite{Kavitha,Akbas1,Mirkin}) have been often adopted for quantifying the similarity between two sets (categorical or binary data), concepts such as the Pearson correlation and cosine similarity have been widely employed for measuring the similarity between two real-valued feature vectors (e.g.  ~\cite{De_Arruda2012,Putra2017,Steinley,Lieve}).

The \emph{coincidence} similarity index has been proposed as an extension of the Jaccard index in the sense of being applicable to real-valued data and being able to reflect the relative interiority between the datasets~\cite{CostaJaccard}.  The extension of the Jaccard index to real-valued data has been accomplished through multiset theory, suitably adapted to take into account negative values~\cite{CostaJaccard,CostaMset,CostaMNeuron}.  The Jaccard index has been shown~\cite{CostaJaccard} to be unable to take into account the relative interiority, or overlap (e.g.~\cite{Kavitha}) between two sets.  This can be addressed by multiplying the Jaccard index with the interiority index, therefore yielding the \emph{coincidence} similarity index.  

The potential of the coincidence index for more strict quantification of the similarity between two generic mathematical structures has been verified respectively to several comparative approaches (e.g.~\cite{CostaComparing,CostaMNeuron}) and applications (e.g.~\cite{Domingues_CityMotifs,Costa_CitiesSimilarity}).

In addition, the real-valued coincidence similarity index can be adapted, with the introduction of a parameter $\alpha$, so as to allow the contributions of pairs of features with the same or opposite signs to be taken with different weights respectively to the resulting coincidence values~\cite{CostaSimilarity,CostaJaccard,CostaMNeuron}.    

A method for translating datasets, with data elements characterized by respective features, into respective detailed networks has been reported~\cite{CostaCCompl} which is based on assigning a node to each of the data elements, while the pairwise links between nodes correspond to the coincidence similarity between the respective standardized (or not) features.  This method has been verified to allow networks characterized by a marked level of interconnection detail as well as by enhanced modularity~\cite{CostaCCompl}.  As discussed in~\cite{CostaMNeuron}, the coincidence index naturally performs the comparison relatively to the magnitudes of the  vectors, being little susceptible to perturbations or noise on single features.

Only two parameters are involved in the coincidence methodology, namely the above mentioned $\alpha$ as well as an overall threshold $T$ which is eventually adopted in case binary connections are required.  By varying $\alpha$ and $T$, the coincidence approach to translating dataset into networks can be optimized respectively to sought properties such as network modularity, number of components, among many alternative possibilities.

In its parameterless version, the coincidence similarity index between two vectors $X$ and $Y$ can be defined as corresponding to the product between the respective Jaccard and interiority indices, namely:
\begin{equation}
  \mathcal{C}(X,Y) = \mathcal{J}(X,Y) \mathcal{I}(X,Y)
\end{equation}

The real-valued coincidence can be immediately obtained as:
\begin{equation}
  \mathcal{C}_R(X,Y) = \mathcal{J}_R(X,Y) \mathcal{I}_R(X,Y)
\end{equation}

The interiority (or overlap, e.g.~\cite{Kavitha}) index, in its real-valued version~\cite{CostaSimilarity}, can be expressed as:
\begin{equation}
  \mathcal{I}_R(X,Y) = \frac{\sum_{i \in S} \min\{|x_i|,|y_i|\}}{\min\{S_X,S_Y\}} 
\end{equation}
where $S$ is the support of the vectors, and
\begin{equation}
  S_X = \sum_{i \in S} |x_i|;  \quad S_Y = \sum_{i \in S} |x_i|
\end{equation}
with $0 \leq \mathcal{I}_R(X,Y) \leq 1$.

The real-valued Jaccard index can be defined~\cite{CostaSimilarity,CostaJaccard} as:
\begin{equation}
  \mathcal{J}_R(X,Y) = \frac{\sum_{i \in S} sign(x_i y_i)  \min\{|x_i|,|y_i|\}}{\sum_{i \in S} \max\{|x_i|,|y_i|\}}
  \label{eq:parameterless}
\end{equation}
with $0 \leq \mathcal{J}_R(X,Y) \leq 1$.

Now, the parametrized real-valued coincidence index can be written as:
\begin{equation}
  \mathcal{J}_R(X,Y,\alpha) = \frac{[\alpha]s_+(X,Y) - [1-\alpha]s_-(X,Y)}{\sum_{i \in S} \max\{|x_i|,|y_i|\}}
\end{equation}
where:
\begin{equation}
  s_+(X,Y) = \sum_{i \in S} |sign(x_i) + sign(y_i)|  \min\{|x_i|,|y_i|\} 
\end{equation}
\begin{equation}
  s_-(X,Y) = \sum_{i \in S} |sign(x_i) - sign(y_i)|  \min\{|x_i|,|y_i|\} 
\end{equation}

The parameter $\alpha$, with $0 \leq \alpha \leq 1$ provides an interesting manner to control the relative contributions of the pairwise features combinations with the same or opposite signs onto the resulting similarity values.  In particular, for $\alpha > 0.5$, the pairwise combinations of features with the same sign will have larger relative contribution to the resulting index than those with opposite signs.  By varying $\alpha$, it becomes possible to optimize the obtained coincidence similarity values with respect to several indices, such as the modularity of the resulting networks~\cite{CostaCCompl}.

It can be verified that when $\alpha=0.5$, the parameterized real valued coincidence becomes identical to its parameterless version.

\subsection{Translating Datasets into Networks by Using the Coincidence Methodology}

As proposed in~\cite{CostaCCompl}, the coincidence index can be effectively employed to translate datasets into complex networks.  Each of the $N$ data elements $x_i$ is characterized by $M$ respective features $j$, $1 \leq j \leq M$, organized into respective \emph{feature vectors}, which are then standardized respectively to each of the features so as to yield null means and unit variance.  The standardization is performed respectively to each feature $j$ of each data element $i$ as:

\begin{equation}
   \tilde{x}_{i,j} = \frac{x_{i,j}-\mu_i}{\sigma_i}
\end{equation}

where $\mu_i$ and $\sigma_i$ are the means and standard deviation along the values of the feature $j$.

The coincidence similarity is then calculated respectively to each pair of data elements, yielding a respective adjacency matrix which can be thresholded by $T$ (the coincidence values of the entries larger or equal to $T$ are kept).  The parameter $\alpha$ can be set so as to optimize some property of interest, such as the network modularity or the number of isolated nodes.

Provided the parameters $\alpha$ and $T$ are set in a suitable manner, this simple methodology has been found~\cite{CostaCCompl,CostaComparing,CostaCaleidoscope,Costa_CitiesSimilarity,Domingues_CityMotifs} to allow networks to be obtained that are characterized by enhanced level of details and modularity.  These enhanced features are a direct consequence not only of the more strict ability of coincidence to quantify similarity~\cite{CostaSimilarity,CostaMNeuron}, but also of the possibility to control, by setting the parameter $\alpha$, the relative contribution of features with the same or opposite sign on the resulting similarity value.

\subsection{Network Optimization}

Conceptually, a \emph{community} or \emph{module} of a network is a set of nodes that are more intensely interconnected with one another than with the remainder of the network~\cite{Newman2004,Clauset2004}.  Networks that have communities, which are said to be \emph{modular}, appear frequently in real-world systems, including social networks, scientific collaboration, as well as several biological networks, including protein interaction and, very probably, also enzyme similarity~\cite{Girvan2002,Lewis2010}.  

In order to quantify how modular a network is given a proposed division criterion, a possibility is to calculate the network \emph{modularity}~\cite{Clauset2004,Newman2004,Brandes2008}.  The modularity quantifies the difference between the number of edges that connect nodes within and among each candidate module.  There are several ways in which the modularity of a network can be defined, but, in this work, we adopt the equation~\cite{Brandes2008}:

\begin{equation}
   \label{eq:modularity}
   \mathcal{Q}(\mathcal{C}) =  \sum_{c\, \in\, \mathcal{C}} \left[\frac{|E(c)|}{m} - \left(\frac{\sum_{v\, \in\, c} deg(v)}{2m}\right)^2 \right]
\end{equation}

Where $-1/2 \leq \mathcal{Q}(\mathcal{C}) \leq 1$, $\mathcal{C}$ is a proposed community structure, $c$ is a module, $m$ is the number of nodes in the network, $E(c)$ is the number of edges within the $c$ module, and $deg(v)$ is the total degree of the vertex $v$~\cite{Brandes2008}.  This simple equation can be used to optimize the modularity of a network: the first term should be maximized with modules that have many edges among their nodes, while the second term should be minimized with the module having nodes with a low degree.  An equivalent and computationally efficient approach to calculate this equation~\cite{Clauset2004} was used in this work.

Although there are several manners of defining communities on a network based on its properties~\cite{Newman2004,Fortunato2010}, the community structure for biological networks usually has a biological meaning transcending its topology.  In the particular cases of enzymes, this has to do with their catalytic activity.  In our case, the enzyme similarity networks will be divided into four modules respective to the four enzyme types under study.  Therefore, this proposed community structure is expected to be preserved meaning that, in this work, we aim to determine suitable interconnections between the network nodes which lead to a significant agreement with the original four modules.  This will be approached by using only the coincidence method while varying the $\alpha$ and $T$ parameters.  In our analysis, the modularity  $\mathcal{Q}(\alpha,T)$ can be seen as a function of these two parameters.

An unwanted consequence of the optimization of the equation~\ref{eq:modularity} is that networks with a low quantity of edges and with multiple isolated nodes may have high values of modularity.  Although this is not necessarily a problem, the proposed coincidence method to create enzyme similarity networks are intended to be used to create modular networks -- thus with biologically relevant information -- with a reduced number of isolated nodes -- thus integrating most of the enzymes under analysis.  

Therefore, we propose to combine the increase of the modularity $\mathcal{Q}(\alpha,T)$ with the minimization of the number of isolated nodes $I$ into a single combined optimization index $\mathcal{F}$~\cite{Audet2008}:

\begin{equation}
   \label{eq:optimization_index}
   \mathcal{F}(\alpha,T) =  (I(\alpha,T)/m - 0)^2 + (\mathcal{Q}(\alpha,T) - 1)^2
\end{equation}

The index $F(\alpha,T)$ makes possible to both minimize the relative number of isolated nodes $I(\alpha,T)/m$ and maximize the modularity $\mathcal{Q}(\alpha,T)$ of the network (considering the biologically relevant community structure $\mathcal{C}$), solely by varying $(\alpha,T)$.

In brief, in the present work, given any set of the 6 physicochemical features under study, we apply the coincidence methodology to create 121 different networks between the 70 adopted model enzymes, each network corresponding to a pair of $(\alpha,T)$, with $\alpha$ and $T$ assuming values in $0.0, 0.1, 0.2, ..., 0.9, 1.0$.  Then, we estimate the optimization index $\mathcal{F}$ for each network considering the division of the nodes into four enzyme types (modules), and we define the optimal network as the one with $(\alpha,T)$ that yielded the minimal calculated value of $\mathcal{F}$.

\subsection{Feature Analysis}

Before real-world or abstract patterns can be \emph{recognized} or \emph{classified}, they need to have their most relevant properties translated into respective mathematical measurements, or \emph{features}.  While there is no definite method for selecting which are the best features relatively to each pattern recognition problem, it becomes of particular relevance to devise means for studying the effect of specific combinations of features into the obtained result(e.g.~\cite{ShapeBook}).  Of particular relevance is the potential impact of different feature choices on the performance of the implemented approaches respectively to each particular dataset.

In the present work, each of the 70 adopted model enzymes is characterized by 6 physicochemical features.  In order to investigate how possible combinations of these features impact the obtained enzyme similarity network, we adopt the coincidence-based approach reported in~\cite{CostaCCompl}.  This approach involves obtaining \emph{optimal similarity networks} for each of the possible feature combinations and then mapping them into nodes of a \emph{feature combination network}, in which the links between any two nodes are determined by the coincidence between the adjacency matrix of the similarity networks they represent.

The feature combination network expresses in a quantitative and effective manner the effect of each feature in the obtained similarity networks.  This type of network has been found to be strongly modular respective to specific sets of features~\cite{CostaCCompl,Costa_CitiesSimilarity,Domingues_CityMotifs,Costa_ElementaryParticles}, so that each of the identified communities of features can be taken as a set of related models for describing the dataset.  A community that is strongly connected within itself indicates particularly homogeneous models, characterized by intense synergy between the features involved.  

In the present work, we consider the optimal enzyme similarity network for each of the possible combinations of the six physicochemical features under study.  In order to provide a direct indication of the effect of the considered feature combinations on the obtained enzyme similarity networks, the respective optimization index is also shown in the resulting feature combination network.

\section{Results and Discussion}

In this section, we derive the optimal enzyme similarity network from the adopted six physicochemical features respectively to the adopted optimization index (which needs to be minimized) and investigate the impact of the combinations between the six features on the resulting networks.

\begin{figure*}
    \centering
    \includegraphics[width=0.85\textwidth]{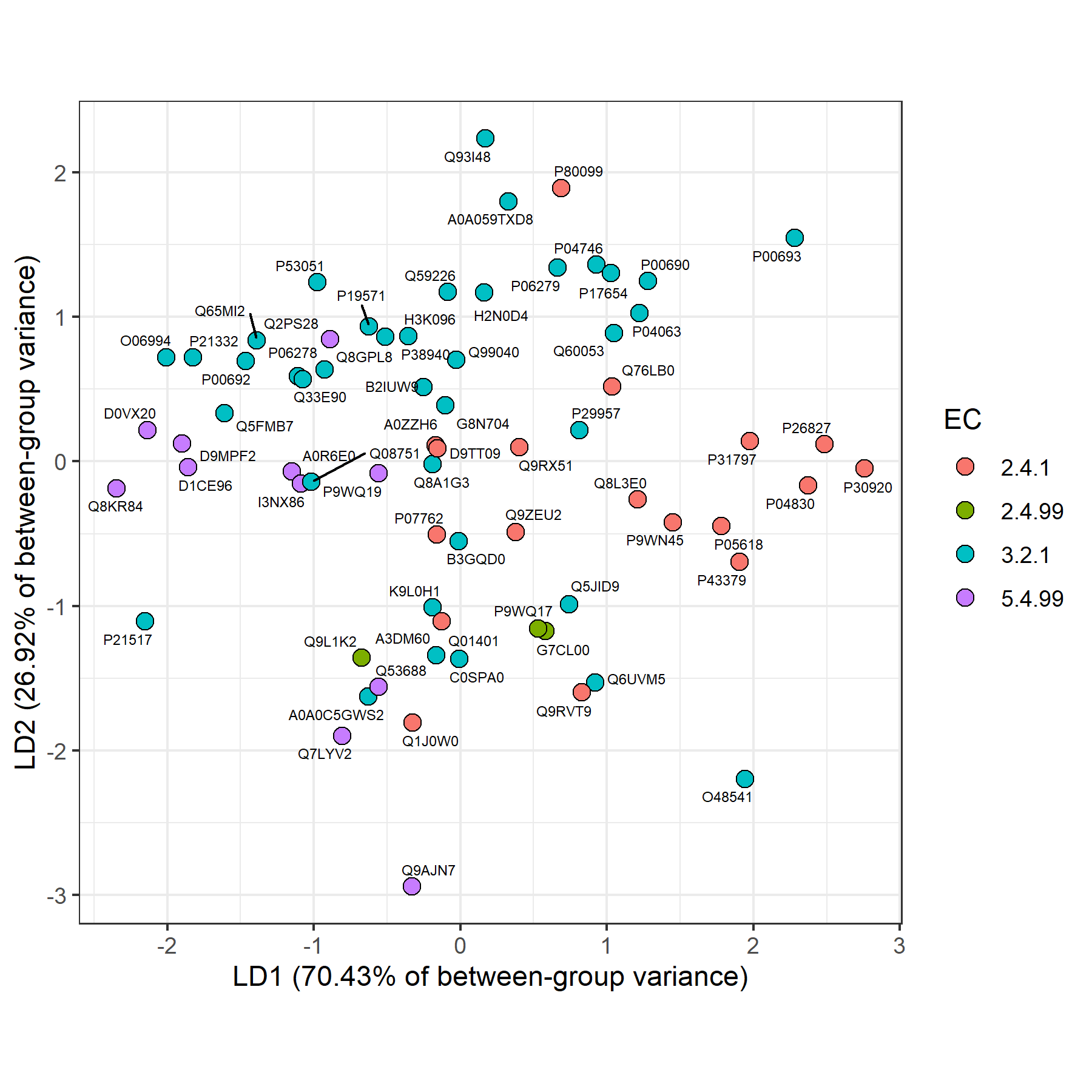}
    \caption{The two-dimensional linear discriminant analysis (LDA) of the 70 model enzymes, derived from their six physicochemical features.  The points are labeled with the respective Uniprot accession numbers of the enzymes.  Little separation between the four enzyme types can be observed from this projection.}
    \label{fig:Biophysical_LDA}
\end{figure*}

We start by presenting, in Figure~\ref{fig:Biophysical_LDA}, the two-axes Linear Discriminant Analysis (LDA, e.g.~\cite{ShapeBook}) projection of the 70 model proteins considering the adopted six physicochemical features.  The LDA consists of a supervised approach in which the original data elements are projected into a lower-dimensional space to maximize their separation based on the respective intra and intercluster distance matrices (e.g.~\cite{ShapeBook}).

It can be readily observed from this result that the LDA projection is unable to effectively cluster the four types of enzymes, being mostly devoid of clustering tendency.  This result is partially due to LDA being a lower-dimensional projection of six-dimensional data, data that represents physicochemical and non-catalytic characteristics of the enzymes.  Therefore, the already expected weak interrelationship between different EC groups in a six-dimensional space may have been weakened by a substantial loss of information.  In addition, LDA is based on correlation-related similarity, which has been found to be less strict than the coincidence approach (e.g.~\cite{CostaComparing,CostaSimilarity,CostaMNeuron}).

\begin{figure*}[h!]
    \centering
    \includegraphics[width=\textwidth]{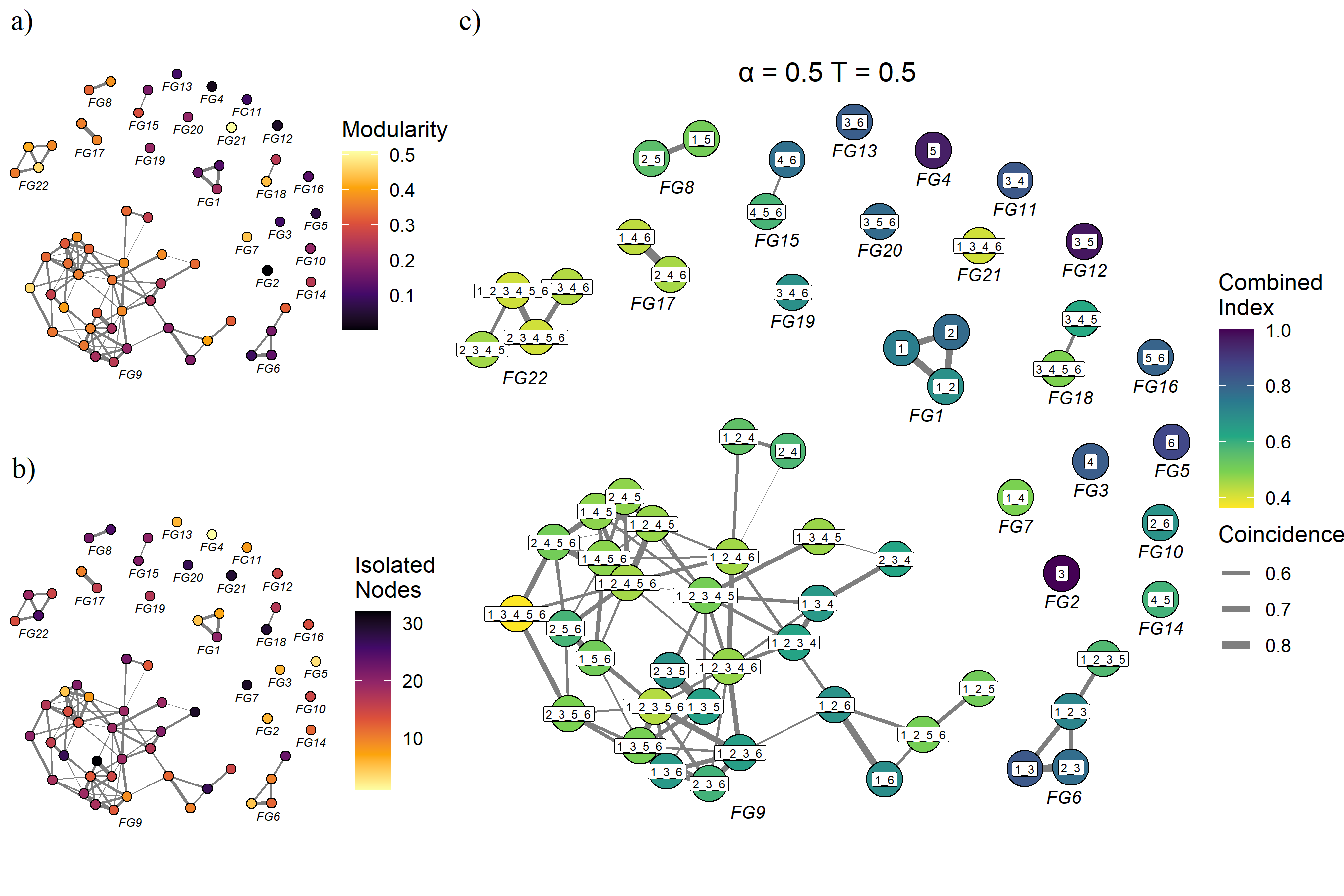}
    \caption{Feature combination network, the network of coincidence similarities between the 63 enzyme networks respective to each possible combination of the physicochemical features.  Each vertex corresponds to an optimal coincidence graph of enzymes created based on the labeled physicochemical features, while the edges represent the coincidence index between pairs of graphs.  The groups of connected vertexes are labeled as Feature Groups (FG).  Each vertex is colored according to the (a) modularity, (b) the number of isolated nodes, and (c) the combined index of their respective enzyme similarity graph.   The label numbers represent the following physicochemical features: 1 - number of amino acids, 2 - molecular weight, 3 - theoretical pI, 4 - instability index, 5 - aliphatic index, and 6 - GRAVY.  This network was obtained for $\alpha = T =0.5$.}
    \label{fig:Biophysical_Plot_Network_of_Networks}
\end{figure*}

Next, we construct 63 optimal \emph{enzyme similarity networks} respectively to the adopted combined optimization index $\mathcal{F}(\alpha,T)$, each one of them respective to different possible combinations of the six physicochemical features under study.  This optimization took place over the two-dimensional space defined by the parameters $\alpha$ and $T$.  The 63 optimal networks were gathered in a \emph{feature combination network} in Figure~\ref{fig:Biophysical_Plot_Network_of_Networks}.  Figure~\ref{fig:Biophysical_Plot_Network_of_Networks}(a) depicts, in respective heatmap colors, the modularity values obtained for each of the 63 combinations of the six physicochemical features.  Figure~\ref{fig:Biophysical_Plot_Network_of_Networks}(b) shows the same feature combination network, but with heatmap colors indicating the resulting number of isolated nodes for each feature combination, and Figure~\ref{fig:Biophysical_Plot_Network_of_Networks}(c) presents the respective combined optimization index values in heatmap colors.  

A large connected component can be readily identified in Figure~\ref{fig:Biophysical_Plot_Network_of_Networks}, corresponding to enzyme similarity networks that are similar to one another, as well as some smaller groups and isolated nodes.   Each of these groups can be understood as corresponding to a possible \emph{model} of the enzyme similarities, in the sense that the nodes within each of these groups will lead to enzyme similarity networks that are similar~\cite{Costa_ElementaryParticles}.  
Distinct connected components, therefore, provide alternative models.  Each of these models can be summarized in terms of the respective hub~\cite{Costa_CitiesSimilarity}.  

The feature groups FG9 are worthy of special attention.  This group corresponds to the largest obtained connected group, composed of combinations encompassing all the six adopted physicochemical features, which also happens to include one of the largest modularity values.  The minimum observed combined optimization index $F$ is to be found within this connected group -- respectively to the features $1$, $3$, $4$, $5$, and $6$.

\begin{figure*}[h!]
    \centering
    \includegraphics[width=0.6\textwidth]{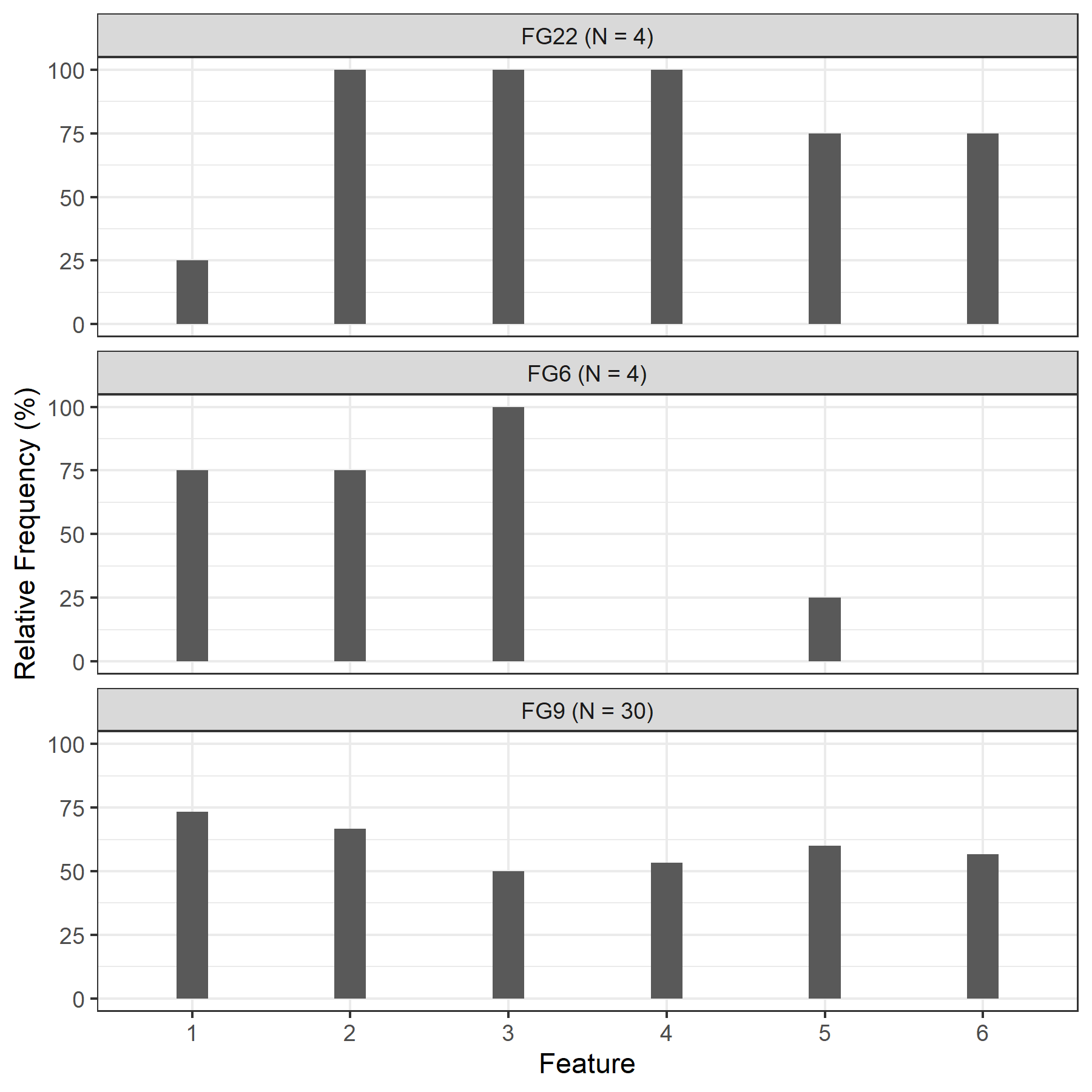}
    \caption{Histograms indicating the frequency of occurrence of each of the adopted six physicochemical features within the main models FG22, FG6, and FG9.  The optimal enzyme similarity network was identified within FG9.}
    \label{fig:Plot_Features_Histogram}
\end{figure*}

Figure~\ref{fig:Plot_Features_Histogram} presents the number of occurrences of each of the six physicochemical features within each of the three largest feature groups, namely FG22, FG6, and FG9.  These histograms indicate that all the six features are almost equally frequent in FG9.  A similar situation, except for a substantially less frequent presence of the first feature, can be observed for FG22, while FG6 contains mostly the three first features as well as one occurrence of the fifth feature.  This shows that these six physicochemical characteristics are mostly interchangeable for creating enzyme networks with similar connections, at least when a minimum of three of them are considered in the calculations.

Table~\ref{tab:max} presents the five feature combinations (nodes in the network of Figure \ref{fig:Biophysical_Plot_Network_of_Networks}) leading to optimal enzyme similarity networks with the lowest observed values of the combined optimization index.  It is presented the $\alpha$ and $T$ values considered to achieve each optimal enzyme similarity network, as well as its respective modularity, number of isolated nodes, number of components, combined index, and feature group (FG).  Except for the fifth entry in this table (features $1$, $4$, and $6$), all other configurations involve at least four out of the six adopted physicochemical features.

\begin{table}[ht!]
\caption{The five feature combinations leading to the most optimized combined indices, indicating the respective parameter configurations ($\alpha$ and $T$), as well as other properties of the respective optimal enzyme similarity networks.  Most of these configurations involve at least 4 of the six adopted physicochemical features.  \label{tab:max}}
\resizebox{\textwidth}{!}{
\begin{tabular}{cc|c|ccccc} \toprule
$\alpha$ & T & Features & Modularity & Isolated Nodes & Components & Combined index F & FG \\ \midrule
0.9 & 1.0 & 1, 3, 4, 5, 6 & 0.4679 & 20 & 35 & 0.3648 & 9 \\
0.5 & 0.6 & 1, 3, 4, 6 & 0.5092 & 29 & 43 & 0.4125 & 21 \\
0.3 & 0.3 & 2, 3, 4, 5, 6 & 0.4659 & 25 & 36 & 0.4128 & 22 \\
0.3 & 0.3 & 1, 2, 3, 4, 5, 6 & 0.4157 & 19 & 32 & 0.4151 & 22 \\
0.5 & 0.6 & 1, 4, 6 & 0.3593 & 10 & 19 & 0.4309 & 17 \\ \bottomrule
\end{tabular}
}
\end{table}

\begin{figure*}[ht!]
    \centering
    \includegraphics[width=\textwidth]{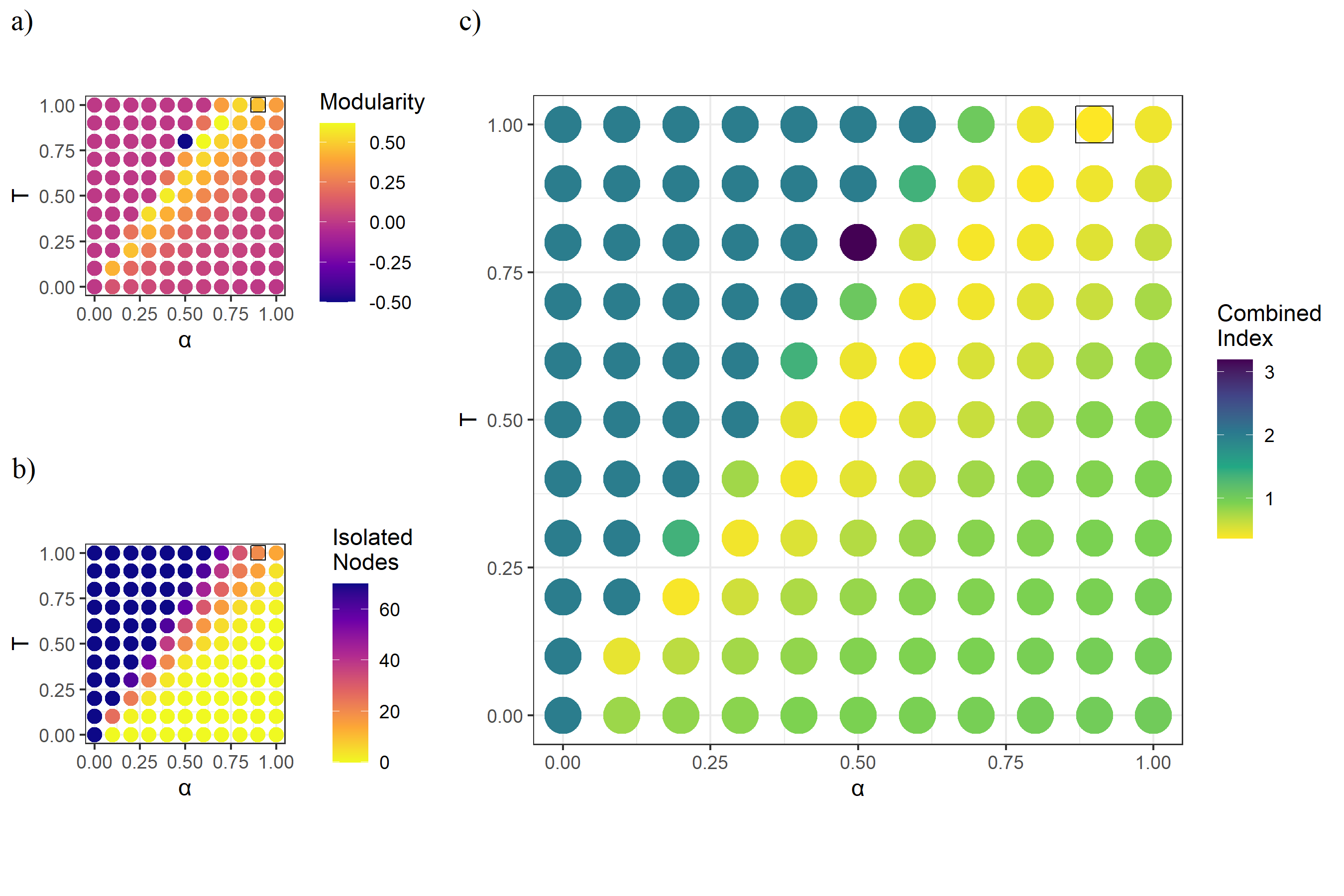}
    \caption{The modularity values (a) obtained for the features $1$, $3$, $4$, $5$, and $6$ in terms of several values of $\alpha$ and $T$.  The respective numbers of isolated nodes are shown in (b), and the combined optimization index values in (c).  This feature combination generates the enzyme similarity network with the lowest value of the combined index observed.  The optimal parameter configuration is obtained for $\alpha=0.9$ and $T=1.0$, indicated by the black box.}
    \label{fig:Biophysical_Optimization}
\end{figure*}

According to the combined optimization index method, the best enzymatic similarity network achieved was constructed considering the features $1$, $3$, $4$, $5$, and $6$, with $\alpha = 0.9$ and $T = 1.0$.  Next, we proceeded to analyze this combination of features now with multiple values of $\alpha$ and $T$, in order to study how the resulting enzymatic network behaves with the alteration of these two parameters.  Figure~\ref{fig:Biophysical_Optimization}(a) depicts the modularity values, in heatmap colors, of the enzyme networks obtained for several values of $\alpha$ and $T$ considering the features $1$, $3$, $4$, $5$, and $6$.  The respective number of isolated nodes are shown in Figure~\ref{fig:Biophysical_Optimization}(b), and the combined optimization index in (c).

Interestingly, the region of largest modularity corresponds to a diagonal ridge extending roughly from the bottom-left to the upper-right portions of the graph in Figure~\ref{fig:Biophysical_Optimization}(a).  This indicates that larger modularity values tend to occur for mutually similar, though not identical, values of $\alpha$ and $T$.   It is worth noticing that this modularity ridge coincides with an abrupt transition (a `cliff') of the number of isolated nodes taking place along the same region in Figure~\ref{fig:Biophysical_Optimization}(b).  This fact tends to imply the highest modularity values to involve an intermediate number of components (orange points in (b)).  The combined optimization index, shown in Figure~\ref{fig:Biophysical_Optimization}(c), corresponds to a smooth ridge.

Of particular relevance is the observation that the optimal configuration was obtained for $\alpha=0.9$, being therefore otherwise impossible to be identified by using the parameterless coincidence index equation (Eq.~\ref{eq:parameterless}), which would keep $\alpha$ fixed ad $0.5$.  This result corroborates the importance of resourcing to the $\alpha$ parameter while translating datasets into respective coincidence networks.

\begin{figure*}
    \centering
    \includegraphics[width=\textwidth]{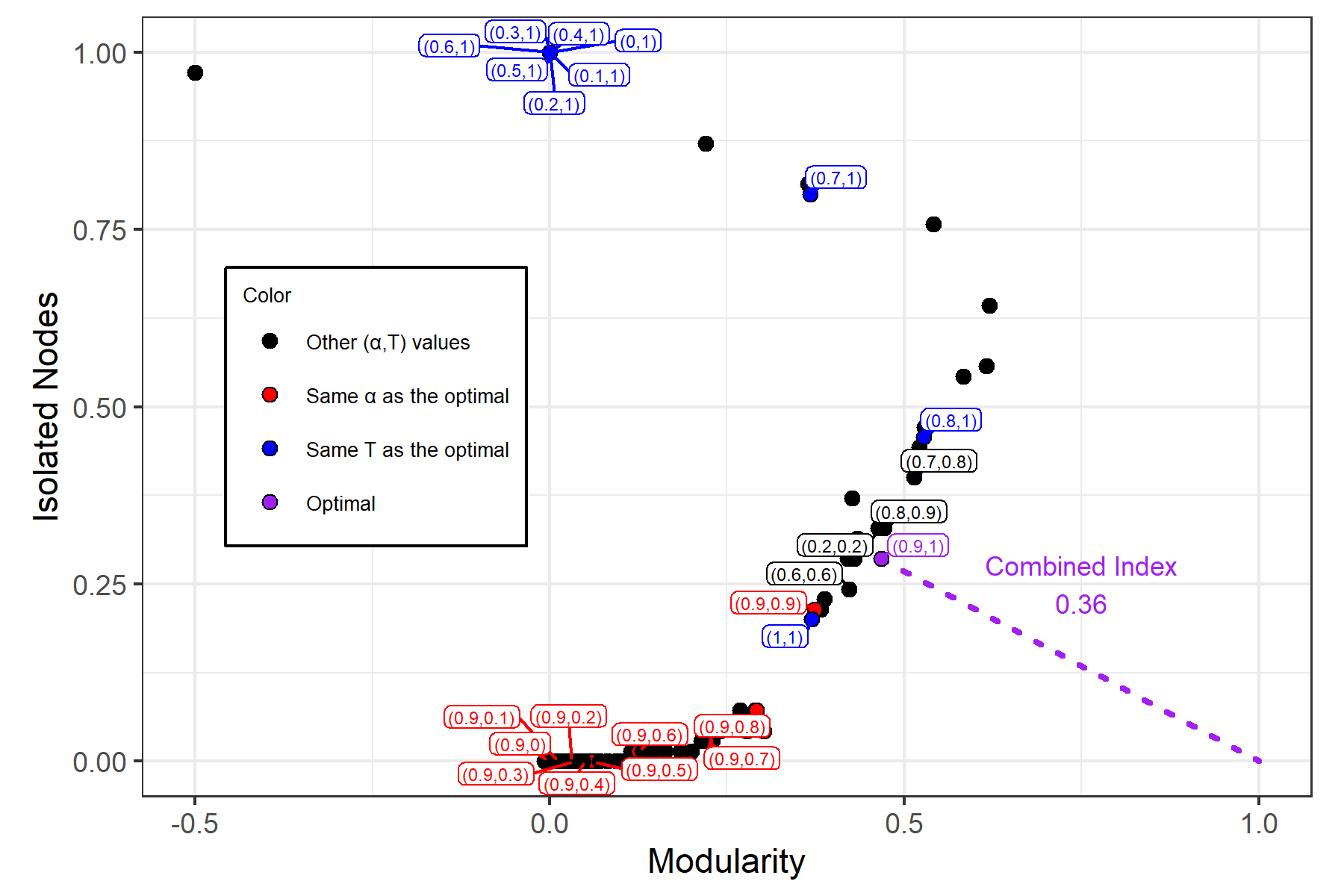}
    \caption{The scatter plot obtained from the modularity and number of isolated nodes of the networks constructed for the features $1$, $3$, $4$, $5$, and $6$ in terms of several values of $\alpha$ and $T$.  The optimal configuration is shown in purple, and those configurations with values of $\alpha$ and $T$ equal to those of the optimal configuration are identified in red and blue, respectively.  The values of these two parameters are shown for several of the combinations in the figure.}
    \label{fig:Biophysical_scatter}
\end{figure*}

Figure~\ref{fig:Biophysical_scatter} shows the scatter plot of the number of isolated nodes in terms of modularity obtained for the enzyme networks of multiple $\alpha$ and $T$ configurations using the features $1$, $3$, $4$, $5$, and $6$.  The optimal combined index configuration is shown in purple, corresponding to $\alpha=0.9$ and $T=1.0$.  Other combinations of special interest are also highlighted in the figure.  It is interesting to observe a relatively well-defined non-linear (V-like) relationship between the two optimization parameters.  

\begin{figure*}
    \centering
    \includegraphics[width=0.8\textwidth]{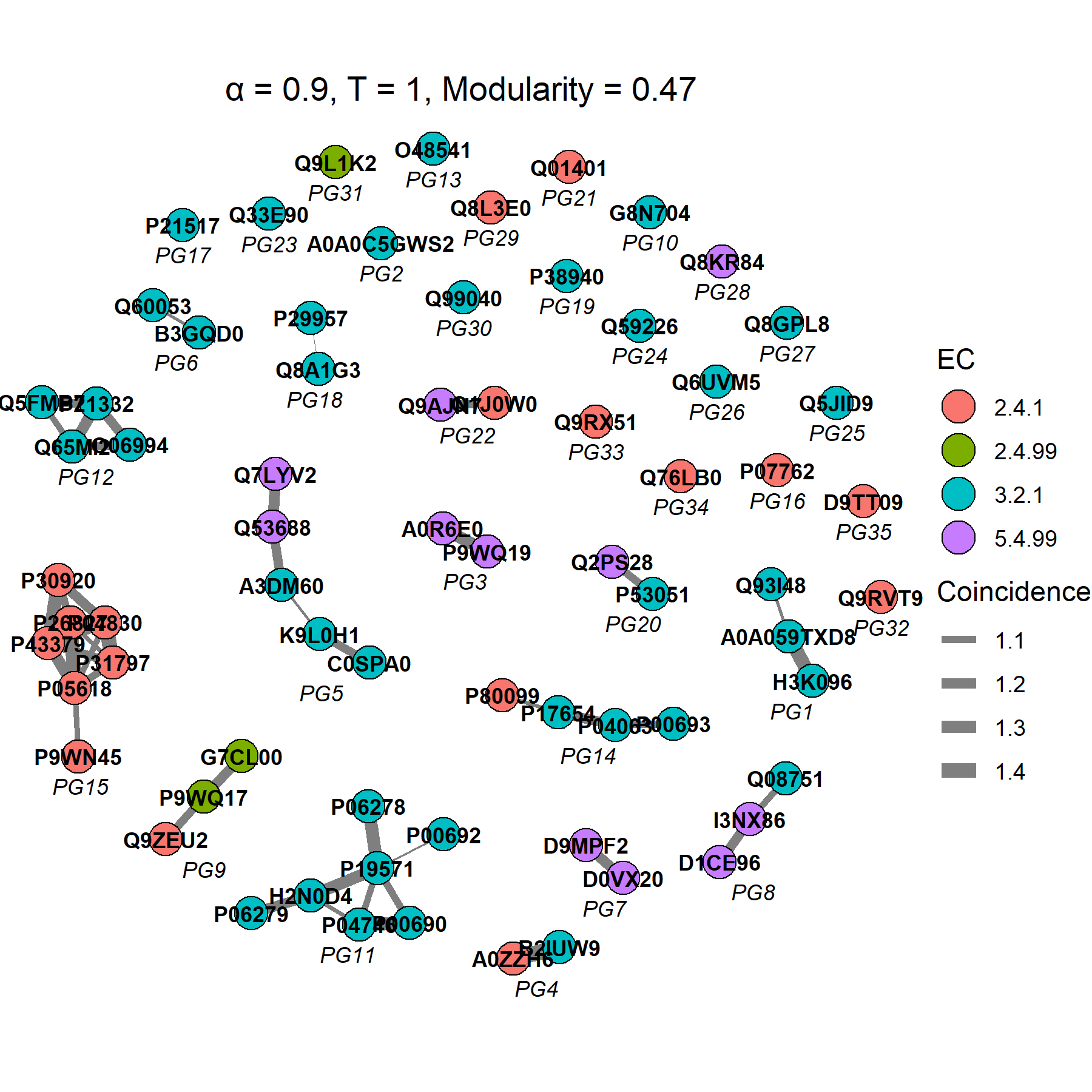}
    \caption{The enzyme similarity network with the optimal combined index, obtained with the features $1$, $3$, $4$, $5$, $6$.  The nodes are labeled with the respective Uniprot accession numbers of the enzymes.  The connected components are labeled as Protein Groups (PG).  Most of these components are strongly interconnected, except for PG5 and PG14, which correspond to chains implied by transitivity of the coincidence pairwise interactions.  Of particular interest is the good homogeneity of enzyme types within each identified connected component, as well as the fact that those four types have yielded separated clusters, suggesting the presence of subcategories within types of enzymes.}
    \label{fig:Best_A_TH_FEATURES}
\end{figure*}

The enzyme similarity network yielding the optimal (minimum) combined index is presented in Figure~\ref{fig:Best_A_TH_FEATURES}.

Several relatively large modules can be identified in  Figure~\ref{fig:Best_A_TH_FEATURES}, most of which are highly uniform in the sense of involving mainly enzymes that catalyze the same type of reaction.  Several of the obtained groups involve strongly interconnected enzymes, indicating respectively high uniformity of physicochemical properties.  Interestingly, some of the obtained communities, such as PG5 and PG14, consist of a string of connected nodes, which therefore indicates transitive similarity between the involved enzymes, with the enzymes corresponding to the two extremities being the least similar among the other enzymes in that group.  Also of particular interest is the fact that all the four enzyme types yielded more than one related connected component, which indicates the presence of subgroups of physicochemical characteristics within the original enzyme types.

From the biological perspective, the relatively good agreement between the obtained groups and the four original enzyme types with the formation of pure type PGs -- such as PG15, PG11, PG12, and PG7 -- substantiate the potential of the physicochemical features $1$, $3$, $4$, $5$, and $6$ in reflecting, as well as providing subsidies for predicting and enhancing, the catalytic activities adopted for the definition of the four original enzyme types. 

Two biochemical explanations can possibly justify the observed relationship between physicochemical properties and enzyme types. First, these five physicochemical features are directly calculated from the amino acids that compose the protein sequences~\cite{ProtParam}, which will also dictate the folding of the enzyme in its final three-dimensional structure and, consequently, its catalytic activity~\cite{Creighton2010}. Second, the physicochemical characteristics of an enzyme are the result of its evolution, which is strongly dependent on its biological function, i.e., its catalytic activities~\cite{Martinez_Cuesta2015}. Although enzymes with similar catalytic activities should not necessarily have similar physicochemical characteristics~\cite{Martinez_Cuesta2015}, the existence of this type of relationship is not unexpected. As shown in~\cite{Mohan2014}, physicochemical properties can be used, with the help of machine learning algorithms, to classify proteins into their correct families.  

\begin{figure*}
    \centering
    \includegraphics[width=\textwidth]{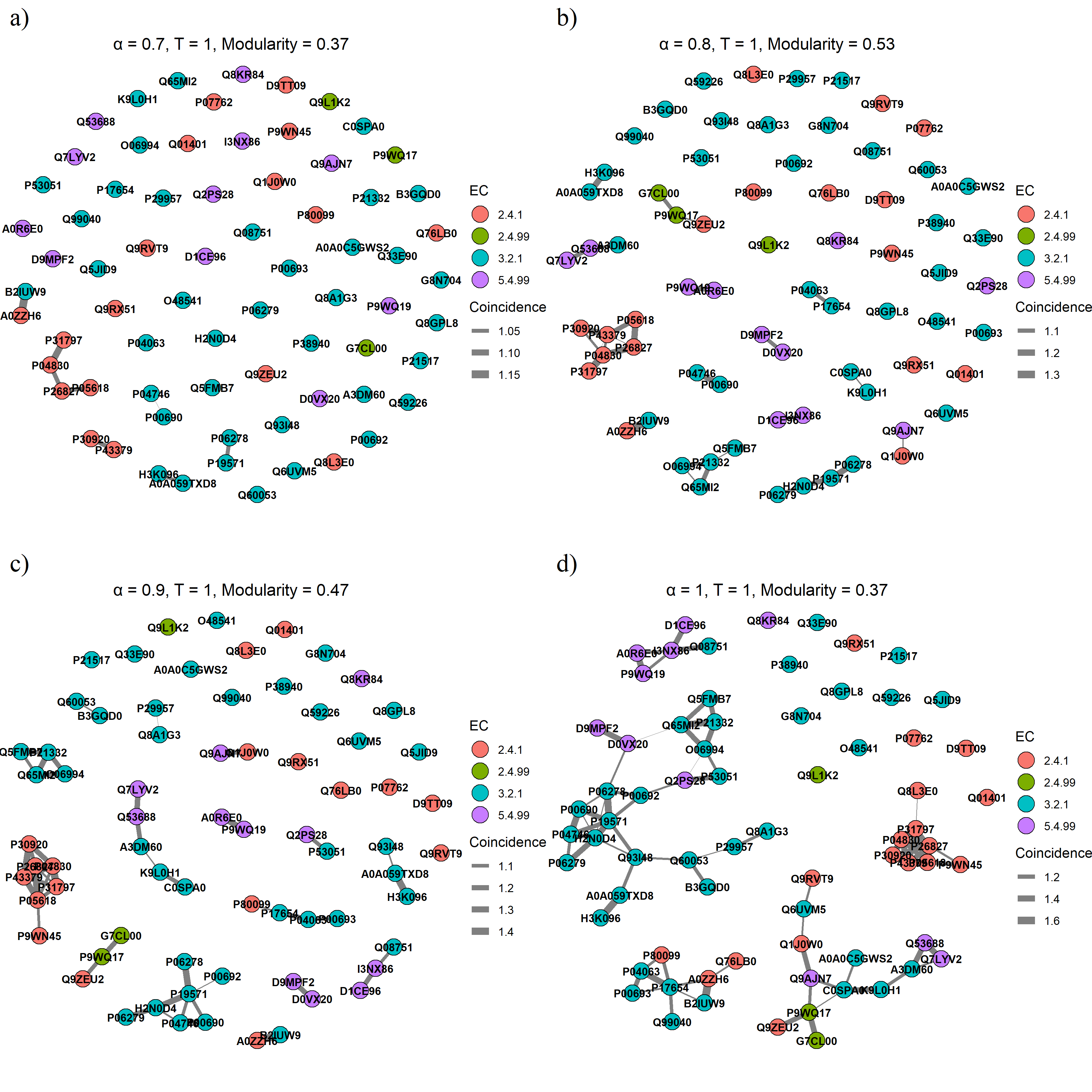}
    \caption{The enzyme similarity networks obtained with the features $1$, $3$, $4$, $5$, and $6$ respectively to four successively larger values of $\alpha$, with $T$ fixed at $1.0$.  The enzymes (nodes) become more and more interconnected as $\alpha$ increases, while the number of components decreases monotonically.  The optimal combined index value considering this set of features is obtained for $\alpha=0.9$ and $T=1.0$.  The width of the links is proportional to the respective coincidence values.}
    \label{fig:Best_A_TH_FEATURES_Multiple_A}
\end{figure*}

Figure~\ref{fig:Best_A_TH_FEATURES_Multiple_A} illustrates the enzyme similarity networks obtained for several equally spaced values of the parameter $\alpha$, more specifically $\alpha = 0.7, 0.8, 0.9, 1$, with $T$ fixed at $1.0$.  The protein similarity network obtained for $\alpha=0.7$ is composed almost exclusively of individual, disconnected nodes, except for a few groups involving from two to four proteins.  As $\alpha$ is progressively increased, the nodes become more and more interconnected, while the combined index reaches its maximum at $\alpha=0.9$, then decreasing for $\alpha=1.0$.  At the same time, the number of connected components decreases monotonically with the increase of $\alpha$.  Even though the combined index values obtained for $\alpha \neq 0.9$ are smaller than the peak value, the respectively obtained networks can still be observed to contain mostly uniform groups of nodes, that are progressively interconnected.  It is interesting to keep in mind that, as $\alpha$ increases, with new links being incorporated into the network, the previous connections are maintained.

\begin{figure*}
    \centering
    \includegraphics[width=\textwidth]{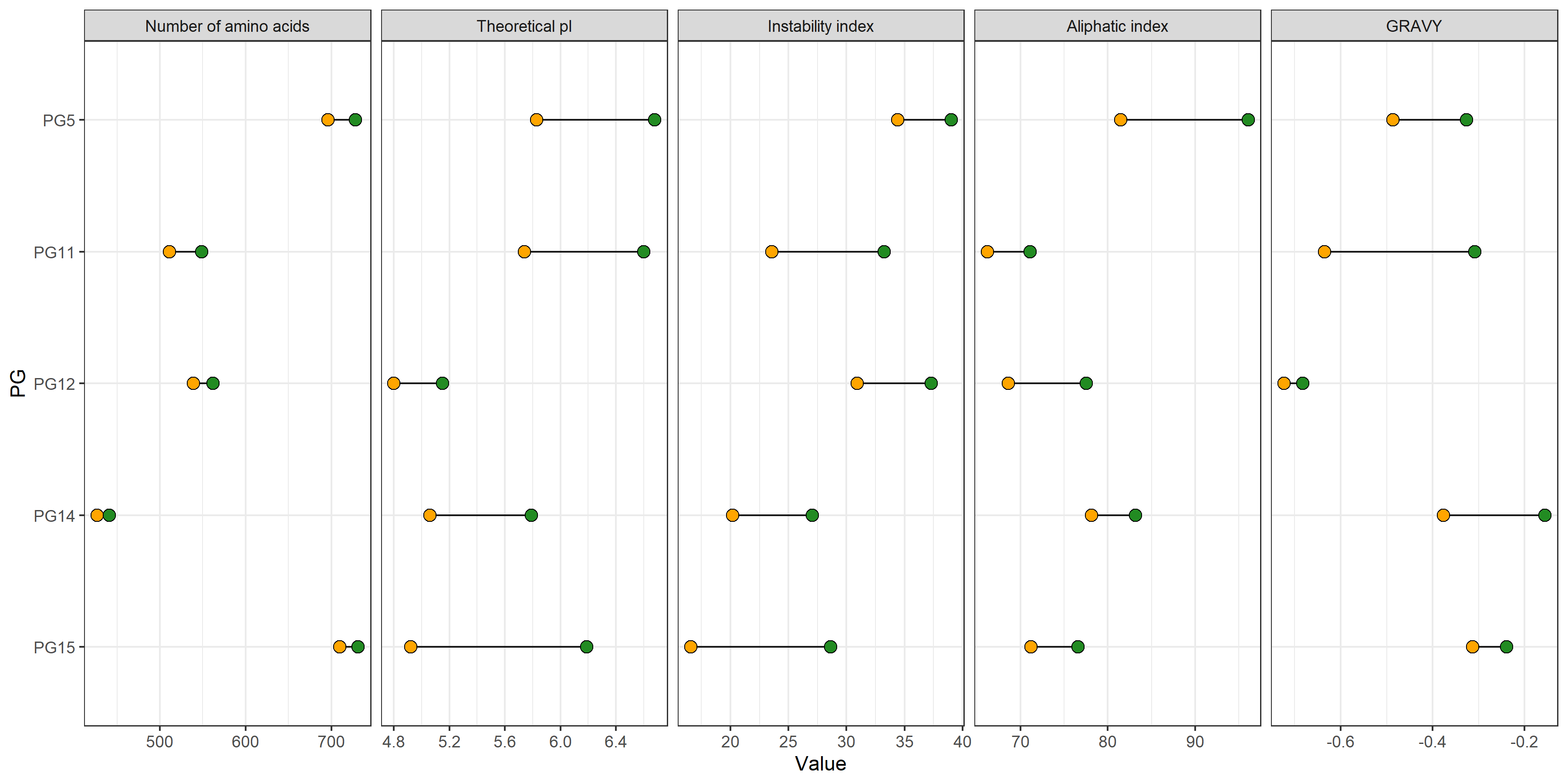}
    \caption{The ranges of values of the five considered physicochemical features observed for the five largest enzymes groups obtained respectively to the optimal enzyme similarity network.}
    \label{fig:ranges}
\end{figure*}

Figure~\ref{fig:ranges} depicts the range of values for each of the five considered physicochemical features characterizing the PGs with at least four enzymes: PG5, PG11, PG12, PG14, and PG15.

According to Figure~\ref{fig:ranges}, the enzymes of each group have quite a similar number of amino acids, thus having similar protein and gene sizes.  Therefore, the enzymes of a given group could be purified with the same size exclusion chromatography equipment, while different groups would require different chromatography gels.  Because the enzymes of the groups PG5 and PG11 have pI values varying from neutral to slightly acidic conditions, these proteins could be purified through both cation- and anion-exchange chromatography, depending on altering the pH of the solutions to acid or basic, respectively.  However, the enzymes of groups PG14, PG12, and, particularly, PG11, because they have lower values of pI, should be more easily purified with anion-exchange chromatography.

Furthermore, by choosing enzymes of the group PG14 or PG15, the biotechnologist will have enzymes with low instability index and high aliphatic index, thus having enzymes with the highest predicted stability for uses in industrial processes.  However, those enzymes might have a slightly hydrophilic behavior, thus they could present problems with solubility.  Alternatively, enzymes of the PG12 group might have more stability problems, but a higher hydrophilic behavior.  Overall, the above analysis illustrates how the proposed methodology could help assemble groups of technically related enzymes for biotechnological uses while taking into account specific enzyme properties and requirements.

\section{Concluding Remarks}

Proteins, which include enzymes, have been largely acknowledged to constitute the building blocks of life.  Their essential role lies in the fact that they implement different types and levels of functionality across species and biological systems, including many functionalities useful for biotechnological applications.  As a consequence, the study of not only the catalytic activities of proteins but their physicochemical properties and applicability outside its original species, as well as how the proteins can be grouped into classes are ubiquitous in biological, biophysical, and biochemical sciences.

The present work has addressed the classification of enzymes based on their physicochemical properties relying on the following three key aspects: (i) the coincidence method is employed to map from several combinations of enzyme features to respective similarity networks; (ii) this mapping is optimized with respect to an optimality index related to the modularity and number of isolated nodes of the resulting networks; and (iii) a coincidence-based approach is applied in order to characterize in a systematic and comprehensive manner the impact of the enzyme features on the resulting optimal similarity networks.  

Several interesting results are reported and discussed, including the identification of  enzyme similarity networks characterized by relatively high modularity levels and small number of isolated nodes obtained from six physicochemical properties of the considered enzymes.  

The several combinations of the six features under study were systematically evaluated, and the optimal enzyme similarity network was obtained while considering features $1$, $3$, $4$, $5$ and $6$, leaving out only the molecular weight.  This result suggests that, among the adopted physicochemical features, the five selected measurements have more pronounced potential for characterizing subgroups of physicochemically related enzymes within the four  types of enzymes.

Another particularly interesting result concerns the importance of using the parameters $\alpha$ and $T$ while optimizing the properties of the networks obtained by the coincidence methodology.  More specifically, it has been shown that the optimal enzyme network could not have been otherwise obtained with the parameterless (i.e., with $\alpha=0.5$) version of the coincidence similarity.  At the same time, the mapping of the modularity and number of isolated nodes in terms of $\alpha$ and $T$ provided interesting insights suggesting that the latter property defines a cliff that nearly coincides with the main diagonal characterized by $\alpha = T$, while the maximum modularity configurations tending to occur along the respectively defined ridge.  Therefore, the maximum modularity tends to take place at positions in the domain $(\alpha,T)$ where the number of isolated nodes undergoes an abrupt variation, which reminds a phase transition.

The described methodology and results pave the way for several interesting prospects for further related studies.  For instance, the reported approach can be directly applied to other important enzyme and protein families, as well as to other types of biological molecules and interactions.  It would also be interesting to consider additional physicochemical and catalytic properties, as well as measurements reflecting the geometrical structure of the considered molecules.  Furthermore, this methodology could be incorporated as a complementary tool in prospecting molecules of biotechnological and pharmaceutical interest or in the rational design of proteins and ligands to improve their physicochemical properties.

\section*{Acknowledgements}

L.  da F.  Costa thanks CNPq (307085/2018-0) and FAPESP (2015/22308-2) for support.  R.  dos Reis thanks CAPES (88887.637015/2021-00) for ﬁnancial support.  This study was financed in part by the Coordenação de Aperfeiçoamento de Pessoal de Nível Superior - Brasil (CAPES) - Finance Code 001

\bibliography{mybib}
\bibliographystyle{unsrt}

\end{document}


\maketitle

\section*{Supplementary Material 1}

\begin{table}[ht!]
\caption{The 70 enzymes from Glycoside Hydrolase Family 13 (GH13) adopted as models in the present work. Data were collected from the CAZy database (http://www.cazy.org/GH13\textunderscore structure.html)~\cite{CAZY}.}

\resizebox{\textwidth}{!}{%
\begin{tabular}{c|cccc}
\toprule
Uniprot Accession Number & Organism & GH13 Subfamily & EC Number & Enzyme Type \\ \midrule
Q76LB0 & Xanthomonas campestris & 23 & 2.4.1.- & 2.4.1 \\
Q01401 & Oryza sativa subsp. japonica & 8 & 2.4.1.18 & 2.4.1 \\
P07762 & Escherichia coli & 9 & 2.4.1.18 & 2.4.1 \\
P9WN45 & Mycobacterium tuberculosis & 9 & 2.4.1.18 & 2.4.1 \\
Q9RX51 & Deinococcus radiodurans & 10 & 2.4.1.18 & 2.4.1 \\
P30920 & Bacillus circulans & 2 & 2.4.1.19 & 2.4.1 \\
P05618 & Bacillus sp. (strain 1011) & 2 & 2.4.1.19 & 2.4.1 \\
Q8L3E0 & Evansella clarkii & 2 & 2.4.1.19 & 2.4.1 \\
P31797 & Geobacillus stearothermophilus & 2 & 2.4.1.19 & 2.4.1 \\
P43379 & Niallia circulans & 2 & 2.4.1.19 & 2.4.1 \\
P04830 & Paenibacillus macerans & 2 & 2.4.1.19 & 2.4.1 \\
P26827 & Thermoanaerobacterium thermosulfurigenes & 2 & 2.4.1.19 & 2.4.1 \\
P80099 & Thermotoga maritima & - & 2.4.1.25 & 2.4.1 \\
D9TT09 & Thermoanaerobacterium thermosaccharolyticum & 18 & 2.4.1.329 & 2.4.1 \\
Q1J0W0 & Deinococcus geothermalis & 4 & 2.4.1.4 & 2.4.1 \\
Q9RVT9 & Deinococcus radiodurans & 4 & 2.4.1.4 & 2.4.1 \\
Q9ZEU2 & Neisseria polysaccharea & 4 & 2.4.1.4 & 2.4.1 \\
A0ZZH6 & Bifidobacterium adolescentis & 18 & 2.4.1.7 & 2.4.1 \\
G7CL00 & Mycolicibacterium thermoresistibile & 3 & 2.4.99.16 & 2.4.99 \\
Q9L1K2 & Streptomyces coelicolor & 3 & 2.4.99.16 & 2.4.99 \\
P9WQ17 & Mycobacterium tuberculosis & 3 & 2.4.99.16 & 2.4.99 \\
P00692 & Bacillus amyloliquefaciens & 5 & 3.2.1.1 & 3.2.1 \\
P06278 & Bacillus licheniformis & 5 & 3.2.1.1 & 3.2.1 \\
Q93I48 & Bacillus sp. KSM-K38 & 5 & 3.2.1.1 & 3.2.1 \\
P06279 & Geobacillus stearothermophilus & 5 & 3.2.1.1 & 3.2.1 \\
P00693 & Hordeum vulgare & 6 & 3.2.1.1 & 3.2.1 \\
P04063 & Hordeum vulgare & 6 & 3.2.1.1 & 3.2.1 \\
P17654 & Oryza sativa subsp. japonica & 6 & 3.2.1.1 & 3.2.1 \\
Q5JID9 & Thermococcus kodakarensis & 20 & 3.2.1.1 & 3.2.1 \\
Q60053 & Thermoactinomyces vulgaris & 21 & 3.2.1.1 & 3.2.1 \\
P04746 & Homo sapiens & 24 & 3.2.1.1 & 3.2.1 \\
H2N0D4 & Oryzias latipes & 24 & 3.2.1.1 & 3.2.1 \\
P00690 & Sus scrofa & 24 & 3.2.1.1 & 3.2.1 \\
B3GQD0 & Bacillus sp. KR-8104 & 28 & 3.2.1.1 & 3.2.1 \\
P29957 & Pseudoalteromonas haloplanktis & 32 & 3.2.1.1 & 3.2.1 \\
Q8A1G3 & Bacteroides thetaiotaomicron & 36 & 3.2.1.1 & 3.2.1 \\
Q8GPL8 & Halothermothrix orenii & 36 & 3.2.1.1 & 3.2.1 \\
A0A059TXD8 & Thermotoga petrophila & 36 & 3.2.1.1 & 3.2.1 \\
G8N704 & Geobacillus thermoleovorans & - & 3.2.1.1 & 3.2.1 \\
P21332 & Bacillus cereus & 31 & 3.2.1.10 & 3.2.1 \\
O06994 & Bacillus subtilis & 31 & 3.2.1.10 & 3.2.1 \\
Q5FMB7 & Lactobacillus acidophilus & 31 & 3.2.1.10 & 3.2.1 \\
P53051 & Saccharomyces cerevisiae & 40 & 3.2.1.10 & 3.2.1 \\
A3DM60 & Staphylothermus marinus & 20 & 3.2.1.133 & 3.2.1 \\
P38940 & Geobacillus stearothermophilus & 20 & 3.2.1.135 & 3.2.1 \\
Q08751 & Thermoactinomyces vulgaris & 20 & 3.2.1.135 & 3.2.1 \\
P21517 & Escherichia coli & 21 & 3.2.1.20 & 3.2.1 \\
H3K096 & Halomonas sp. H11 & 23 & 3.2.1.20 & 3.2.1 \\
Q33E90 & Geobacillus sp. HTA-462 & 31 & 3.2.1.20 & 3.2.1 \\
O48541 & Hordeum vulgare & 13 & 3.2.1.41 & 3.2.1 \\
K9L0H1 & Anoxybacillus sp. LM18-11 & 14 & 3.2.1.41 & 3.2.1 \\
C0SPA0 & Bacillus subtilis & 14 & 3.2.1.41 & 3.2.1 \\
A0A0C5GWS2 & Paenibacillus barengoltzii & 14 & 3.2.1.41 & 3.2.1 \\
B2IUW9 & Nostoc punctiforme & 20 & 3.2.1.41 & 3.2.1 \\
Q6UVM5 & Xanthomonas campestris & 4 & 3.2.1.48 & 3.2.1 \\
Q59226 & Bacillus sp. & 20 & 3.2.1.54 & 3.2.1 \\
Q99040 & Streptococcus mutans & 31 & 3.2.1.70 & 3.2.1 \\
Q65MI2 & Bacillus licheniformis & 29 & 3.2.1.93 & 3.2.1 \\
P19571 & Bacillus sp. (strain 707) & 5 & 3.2.1.98 & 3.2.1 \\
D9MPF2 & Erwinia rhapontici & 31 & 5.4.99.11 & 5.4.99 \\
Q8KR84 & Klebsiella sp. LX3 & 31 & 5.4.99.11 & 5.4.99 \\
Q2PS28 & Paraburkholderia acidophila & 31 & 5.4.99.11 & 5.4.99 \\
D0VX20 & Serratia plymuthica & 31 & 5.4.99.11 & 5.4.99 \\
Q9AJN7 & Arthrobacter ramosus & 26 & 5.4.99.15 & 5.4.99 \\
Q7LYV2 & Saccharolobus shibatae & 26 & 5.4.99.15 & 5.4.99 \\
Q53688 & Sulfolobus acidocaldarius & 26 & 5.4.99.15 & 5.4.99 \\
I3NX86 & Deinococcus radiodurans & 16 & 5.4.99.16 & 5.4.99 \\
P9WQ19 & Mycobacterium tuberculosis & 16 & 5.4.99.16 & 5.4.99 \\
A0R6E0 & Mycolicibacterium smegmatis & 16 & 5.4.99.16 & 5.4.99 \\
D1CE96 & Thermobaculum terrenum & 16 & 5.4.99.16 & 5.4.99 \\ \bottomrule
\end{tabular}%
}
\end{table}

\bibliography{mybib}
\bibliographystyle{unsrt}